\documentclass[twocolumn,aps,prba,superscriptaddress]{revtex4-1}
\usepackage[utf8]{inputenc}
\usepackage{amsmath}
\usepackage{graphicx}
\usepackage{booktabs}
\usepackage{siunitx}
\usepackage[format=plain]{caption}
\usepackage{subfig}
\usepackage[margin=0.9in]{geometry}
\usepackage{textcomp}
\usepackage{gensymb}
\usepackage{sidecap}
\usepackage[english]{babel}
\usepackage[autostyle]{csquotes}
\usepackage{braket}
\usepackage{array}
\usepackage{color}
\usepackage{multirow}

\begin{document}
%\justifying
\captionsetup{justification=Justified,}

\title{Unveiling the crystallization kinetics in Ge-rich Ge$_x$Te alloys by large scale simulations with a machine-learned interatomic potential}

\author{Dario Baratella}
\author{Omar Abou El Kheir} 
\author{Marco Bernasconi$^*$}
\affiliation{Department of Materials Science, University of Milano-Bicocca, Via R. Cozzi 55, I-20125 Milano, Italy}

\begin{abstract}
A machine-learned interatomic potential for Ge-rich Ge$_x$Te alloys has been developed aiming at uncovering the kinetics of phase separation and crystallization in these materials.  
The results are of interest for the operation of embedded phase change memories which exploits Ge-enrichment of GeSbTe alloys to raise the crystallization temperature.
The potential is generated by fitting a large database of energies and forces computed within Density Functional Theory with the neural network scheme implemented in the DeePMD-kit package. The potential is highly accurate and suitable to describe  the structural and dynamical properties of the liquid, amorphous and crystalline phases of the wide range of compositions from pure Ge and stoichiometric GeTe to the Ge-rich Ge$_2$Te alloy.
Large scale molecular dynamics simulations revealed  a crystallization mechanism which depends on temperature. At 600 K,  segregation of most of Ge in excess occurs on the ns time scale  followed by crystallization of nearly stoichiometric GeTe regions. At 500 K, nucleation of crystalline GeTe occurs before phase separation, followed by a slow crystal growth due to the concurrent expulsion of Ge in excess.
\end{abstract}

\maketitle
\footnotetext{Email Address:$^*$marco.bernasconi@unimib.it}
\section{Introduction}

Chalcogenide GeSbTe (GST) phase change alloys are employed in emerging new electronic non-volatile memories called Phase Change Memories (PCMs) \cite{wuttig,noe2017phase,zhang2019designing,pirovano}. These devices feature a rapid and reversible transition of the active material between the crystalline and amorphous phases upon  Joule heating. The two phases correspond to the two states of the memory  which can be discriminated thanks to a large  contrast in the electrical resistivity. Readout of the memory consists of the measurement of the electrical resistance at low bias, while programming is achieved by applying current pulses to amorphize the crystal via melting (reset) or to recrystallize the amorphous phase (set). 

PCMs based on the flagship compound Ge$_2$Sb$_2$Te$_5$ (GST225), which is a pseudobinary alloy on the GeTe-Sb$_2$Te$_3$ tie line, was  brought to market in 2017 by Intel and Micron in a 3D cross-bar architecture (3D XPoint$^{TM}$) \cite{OPTANE2017}. More recently, PCMs have also gained attention for embedded applications, in particular for the automotive sector \cite{Cappelletti2020,redaelli2022}. 
The crystallization temperature T$_x$ of the most commonly used GST225  (420-440 K) \cite{NOE2016}  is, however, too low for applications in embedded memories that require data retention at higher temperatures. 
In fact, a compulsory fabrication step of embedded memories is the soldering process in which the device is exposed to 530 K for a few minutes after the microcontroller code was written in the memory \cite{redaelli2022}.
To meet this requirement, various  materials alternative to GST225 have been explored, including InSbTe, GaSbTe, and InGeTe alloys \cite{kim2007change,kim2009investigation,cheng2011crystallization}. Additionally, doping with N or O atoms \cite{morikawa2007doped} and enrichment with Sb and Ge have been proposed \cite{cheng}. Among these options, Ge-rich GeSbTe alloys emerged recently as the most promising materials with T$_x$ exceeding  600 K for Ge-rich alloys on the Ge-Sb$_2$Te$_3$ tie-line 
 \cite{ZULIANI1,ZULIANI201527,palumbo2017,navarro2013trade}.
 Several  high density embedded memories for microcontrollers have been since then reported in literature
 \cite{sousa2015operation,arnaud2020,kiouseloglou,asild}.

The raise in T$_x$ of Ge-rich GST alloys is ascribed  to the segregation of Ge atoms and phase separation into pure Ge and Ge-poor GST alloy during the crystallization process \cite{ZULIANI201527}. The crystallization is supposed to be slowed down by the mass transport involved in the phase separation, resulting in longer  times for crystal nucleation. Evidences of the phase separation were reported  in both the set process \cite{ZULIANI201527,agati,luong2021some,rahier,privitera2018atomic,privitera2020crystallization,Prazakova,fattorini,cecchi2022crystallization} and  the forming operation of the memory (initialization by the first programming pulse) \cite{palumbo2017}. However, the inhomogeneity due to phase separation  could cause a high cell-to-cell variability. Moreover,  a drift in the electrical resistance with time was reported in both the set and reset states. A resistance drift in the reset state (amorphous phase) is common to all phase change materials as it is ascribed to structural relaxation towards a more stable amorphous structure. The resistance drift in the set state (crystalline), instead, is peculiar to Ge-rich GST alloys as it is due  to recrystallization of a residual amorphous region close to the bottom electrode in the phase separated system \cite{laurin}. Gaining a deeper understanding of the segregation and crystallization processes is thus mandatory to mitigate these detrimental effects.

Several details of the overall process are, however, unclear.
Although several works have revealed the presence of amorphous Ge and of a cubic crystalline phase of GST
\cite{ZULIANI1,agati,luong2021some,rahier,privitera2018atomic,privitera2020crystallization,Prazakova,fattorini,cecchi2022crystallization}, the composition of the cubic phase is unclear because analytical tools such as electron energy loss spectroscopy  or energy dispersive x-ray measurements provide only a composition averaged over different grains including pure Ge grains. The presence of crystalline GeTe 
\cite{rahier} and of Sb-rich alloys \cite{petroni1,petroni2}
was also detected.

 In this respect, atomistic simulations can provide useful insights on both the crystallization and segregation processes. 
 Simulations based on Density Functional Theory (DFT), for instance, revealed that the amorphous phase of GST alloys is unstable with respect to Ge segregation for Ge content above 50 $\%$ \cite{sun2021ab}. A high throughput DFT study also revealed that off-stoichiometric compositions (off the GeTe-Sb$_2$Te$_3$ tie-line) might crystallize in the cubic phase at the operation conditions of the memories \cite{OmarHigh}. These DFT works, however, have only addressed the thermodynamics of the phase transformation and not its kinetics because of the limitations in the time and length scales of DFT simulations. 

 A route to overcome the limitations of DFT methods is the development of machine-learned interatomic potentials generated   by fitting a database of DFT energies and forces. In the past, an interatomic potential for  GeTe \cite{SossoNN} was generated within the Neural Network (NN) scheme proposed by Behler and Parrinello \cite{Behler}, while more recently interatomic potentials for stoichiometric GST225 \cite{kheir2023unraveling} were generated with the NN scheme implemented in the DeePMD code \cite{{DeePMD4,DeePMD2,DeePMD3}} and with the Gaussian Approximation Potential \cite{DeringerNatureEl}. 
 
In the perspective of developing an interatomic potential for large scale simulations of Ge-rich GST alloys, we here address the study of the simpler  Ge-rich Ge$_x$Te binary alloy that shares several properties with the ternary GST system and for which detailed experimental data are available from time resolved reflectivity, x-ray diffraction, transmission electron microscopy and Raman spectroscopy
\cite{carria,gourvest,raoux}. Integration of Ge-rich Ge$_x$Te alloys in PCMs was also reported in Ref. \cite{navarro2013SSE}. We considered in particular the composition Ge$_2$Te which is close to the Ge$_{63}$Te$_{37}$ alloys studied experimentally  in Ref. \cite{carria}.

We developed a NN potential suitable to describe Ge$_2$Te and the products of its crystallization process, stoichiometric GeTe and pure Ge, by using the DeePMD code \cite{DeePMD4,DeePMD2,DeePMD3} already employed in our recent work on GST225 \cite{kheir2023unraveling}.
We validated the potential by analyzing the structural  and dynamical properties of Ge$_2$Te, GeTe, and Ge  in the crystalline, liquid, and amorphous phases. Then, we exploited the potential to perform large-scale simulations of the crystallization and phase separation processes.

\section{Methods}

\noindent
We generated the NN potential for  Ge-rich Ge$_x$Te alloys by fitting a database of DFT energies, forces, and virial tensors of 115000  atomic configurations of Ge, GeTe and Ge$_2$Te in the amorphous, liquid and crystalline phases within the framework implemented in the DeePMD code \cite{{DeePMD4,DeePMD2,DeePMD3}}.  Amorphous and liquid Ge, GeTe and Ge$_2$Te were modeled in cells of 144, 108 and 100 atoms (composition Ge$_{67}$Te$_{33}$), respectively. The crystalline phases of Ge and GeTe were modeled in cells with 64 and 96 atoms. The atomic configurations were extracted from DFT molecular dynamics (MD) simulations by using the CP2k code \cite{quickstep2}. Kohn-Sham orbitals were expanded in a Triple-Zeta-Valence-plus-Polarization (TZVP) basis set while the electronic density was expanded in plane waves up to a kinetic cutoff of 100 Ry. The Brillouin Zone (BZ) integration was restricted to the $\Gamma$-point in all MD simulations. The
time step was set to 2 fs and
configurations were extracted every 100-200 fs. The number of configurations for each phase and composition is given in Table \ref{database}. We recalculated energies, forces, and stresses  for each configuration added to the database with a higher accuracy by increasing the kinetic energy cutoff for the plane-waves expansion of the electronic density to 400 Ry and by using a 4x4x4 k-point mesh for the BZ integration.  A Fermi-Dirac smearing in the occupation of Kohn-Sham states was used with an electronic temperature of 300 K.

\begin{table}
\centering
\begin{tabular}{lccccc}
\hline
Phase & Number of configurations \\
\hline
  Ge   {\rm l \& a}   & 27000\\
 GeTe   {\rm l \& a}   & 35000 \\ 			
  Ge$_2$Te {\rm l \& a} & 46000 \\
Crystalline Ge   & 3400 \\ 
Crystalline GeTe & 3600 \\
\hline
\end{tabular}
\caption{Number of configurations included in the database for each composition and phase. {\rm l \& a}   stands for liquid and amorphous phases.}
\label{database}
\end{table}

In NN schemes for the generation of interatomic potentials, the total energy of the system is written as the sum of individual atomic energies that depend on the local environment of each atom.
In the DeePMD scheme the local environment is encoded by local descriptors which are generated by an embedded neural network. A second neural network  (fitting network)  is built for the calculation of energies and forces with the local descriptors as input layer. We designed the embedded network with three hidden layers of 40, 80, and 160 nodes. The cutoff radius r$_c$ was set to 7 \r{A}, which is beyond the third coordination shell of our systems, while the smoothing radius r$_s$ was set to 2 \r{A} \cite{DeePMD2}. The maximum number of neighbors was set to 80. 
In the embedded network, we also exploited the attention mechanism that was recently implemented in the DeePMD code \cite{DeePMD2}.
Finally, the network for the fitting of energy and forces consists of 3 hidden layers with 320 nodes each.
In the embedding and fitting network, we have used
the hyperbolic tangent as an activation function.

The NN potential was generated in an iterative manner. A first version of the potential was generated by using a small training database of about 6000 configurations of Ge, GeTe and Ge$_2$Te.  Then, the potential is used to perform MD simulations that provide high-energy configurations to enrich the training database for a second generation of the potential and at the same time we enriched the database with additional atomic configurations extracted from DFT-MD simulations at different densities and exploring a wider temperature range. We iterated this procedure until the potential reached a satisfactory level of accuracy.

Structural and dynamical properties obtained from NN simulations were  compared with results from DFT-MD simulations  performed with the CP2k code with the same parameters given above. NN-MD simulations  were performed by using the LAMMPS code as MD driver with the DeePMD plugin \cite{LAMMPS}.

To assess whether an atom is crystalline in the simulations of the crystallization process,  we used the Steinhardt order parameter Q$^{dot}_{n}$ \cite{q4}. In general, the Q$^{dot}_{n}$  parameter of order $n$  is defined for each atom $i$  by
\begin{equation}
     Q^{dot}_{n}=\frac{1}{N_i}\sqrt{\sum_{j=1}^{N_i} \sum_{m=-n}^{n}q_{nm,i} q^*_{nm,j}} 
\end{equation}

\begin{equation}
 q_{nm,i}=\frac{1}{N_i} \sum_{j=1}^{N_i}Y_{nm}\left(\mathbf{\hat{r}}_{ij}\right),
\end{equation}

where  
$N_i$ in the number of neighbors of atom $i$ up to a given cutoff, $j$ is the neighbors index, $Y_{nm}$ is the $n$ order spherical harmonic with degree $m$, and $ \mathbf{\hat{r}}_{ij}$ is the unit vector connecting the two atoms.  For the crystallization of GeTe, we considered the Q$^{dot}_{4}$ order parameter, but also the Q$^{dot}_{6}$ has been considered for some analysis as it will be discussed in the relevant section.
The distribution of Q$^{dot}_{4}$ in crystalline and amorphous phases of GeTe, shown in Fig. S1 in the Supplementary Information, suggests  a threshold  of Q$^{dot}_{4}$ = 0.87 for an atom to be crystalline. 

To assess whether a Ge atom is segregated in  regions of amorphous Ge (a-Ge), we calculated the SOAP (Smooth Overlap of Atomic Position) similarity kernel $k_j$ for each atom \cite{gap}. $k_j$ quantifies the similarity of the atomic environment around atom $j$  with a reference atomic environment which is taken here as the average environment in a-Ge at 600 K. $k_j$ is a number that ranges from 0, when the atomic environment is totally different from that of the reference system, to 1 when it is identical. In the SOAP formalism the local atomic density around each atom $j$  is expressed as a sum of Gaussian functions (here with broadening $\sigma_{at}$= 0.3 \r{A}) centered on the position of its  neighbors
up to a given cutoff (9 \r{A} here). Then the density around atom $j$ is expanded in spherical harmonics and radial basis functions $g_{b}(|\mathbf{r}|)$ as
$    \rho_j (\mathbf{r}) = \sum_{blm} c_{blm} g_{b}(|\mathbf{r}|) Y_{lm}(\mathbf{\hat{r}})$.
The coefficients of this expansion define  the so-called power spectrum matrix 
$    p(j)_{b_1 b_2 l} = \pi \sqrt{8/(2l+1)}\sum_m (c_{b_1lm})^*  c_{b_2lm}$
whose elements are turned into a vector $\mathbf{p}_j$ from which
 the SOAP kernel k$_j$ is calculated as
$    k_j = (\mathbf{p}_j/|\mathbf{p}_j|\cdot\mathbf{p}_{ref}/|\mathbf{p}_{ref}|)^{\xi}$,
where $\mathbf{p}_{ref}$ is the average power spectrum of the atoms in the atomic environment of the reference system and $\xi$ is an integer set to 2  in our case. The SOAP similarity kernel has been used in a variety of studies to analyze the atomic structure \cite{bartokcomp,bartokphys} and very recently to discriminate between the crystalline and amorphous/liquid phases in GST alloys \cite{mazza124}. {In this work, we calculated the SOAP kernel using the DScribe Python package \cite{dscribe2,dscribe} within the ASE Python library \cite{ase}.}

\section{Results and Discussion}

\subsection{Validation of the neural network potential}

\begin{figure*}[htbp!]
    \centering
    \includegraphics[scale= 0.8]{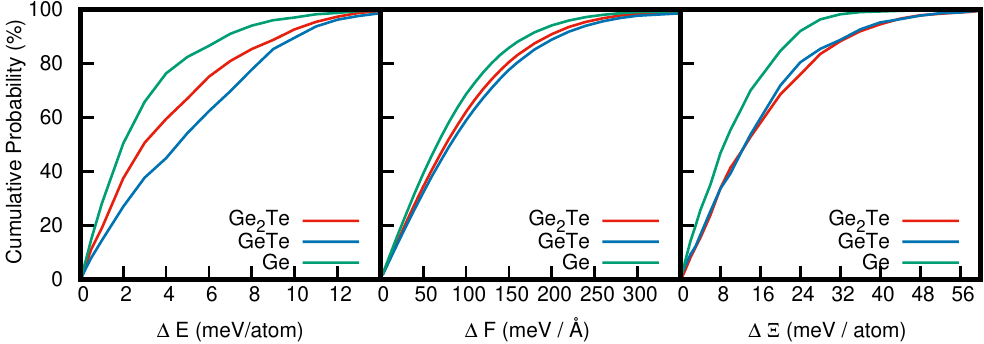}
    \caption{Cumulative fraction of error on energies, forces and virial of the NN potential in Ge, GeTe and Ge$_2$Te.}
    \label{Error}
\end{figure*}

 The accuracy of the NN potential is assessed by the cumulative distribution of the error on energies, forces and virial tensors shown in Fig. \ref{Error}  for Ge$_2$Te, GeTe and Ge. For 80\% of the configurations, the error on the energies, forces and virial is smaller than 8 meV/atom, 150 meV/\r{A} and  25 meV/atom. In particular, among our systems, GeTe has the largest errors, while Ge$_2$Te and Ge have significantly lower errors. Overall, the root mean square error (RMSE) on the energy is 4.4 meV/atom, on forces is 105 meV/\r{A} and   on virial is 14.5 meV/atom.
We remark that the typical average  error obtained with DeePMD for highly disordered phases of multicomponent systems like ours (i.e  liquid and/or amorphous phases)  are in the range 2-7 meV/atom and 90-145 meV/\AA 
 \cite{RMSE1,RMSE2,RMSE3,RMSE4}.

The NN potential has then been validated on the properties of the liquid, amorphous and crystalline phases as described in the separate sections below for pure Ge, stoichiometric GeTe and Ge$_2$Te. 

\subsubsection{The liquid phase}
\begin{figure*}[htbp!]
    \centering
    \includegraphics[scale= 0.8]{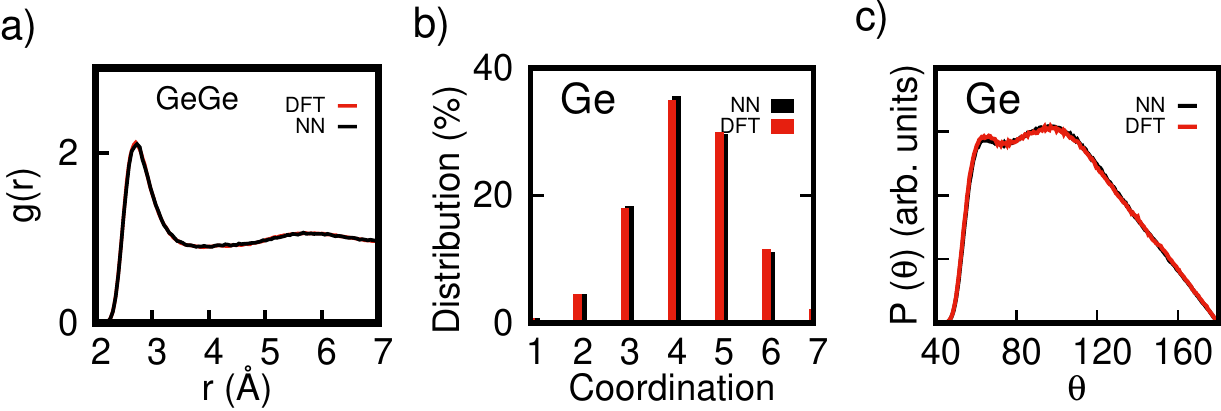}
    \caption{Structural properties of liquid Ge at 1250 K from NN-MD (black lines,  2400 atoms) and DFT-MD (red lines,  300 atoms). The experimental density of the amorphous phase  of 0.0438 atom/\AA$^3$ is used. a) Pair correlation function. b) Distribution of the coordination numbers by assuming a bonding cutoff of 3 \r{A}. 
    Ge is primarily 4-fold and 5-fold coordinated, with a minority of 3-fold and 6-fold coordinated atoms. 
    The average coordination number of Ge is 4.28 with the NN potential and 4.31 with DFT. c) Bond angle distribution function. The peak around 100$^\circ$ is a feature of the Ge atoms in a tetrahedral environment, while the peak at 60$^\circ$ highlights overcoordinated 5-fold and 6-fold atoms. }
    \label{GR_GE_LIQ}
\end{figure*}

We computed the structural properties of liquid Ge at  1250 K  from NN-MD simulations in a 2400-atom model and from DFT-MD simulations  in a 300-atom model. In the perspective to study later the amorphous phase, we used for the liquid the experimental density of the amorphous phase  of 0.0438 atom/\AA$^3$ \cite{baribeau}. The models were equilibrated first at 2000 K for 10 ps  and then at 1250 K for 20 ps. The structural properties were evaluated from the last 10 ps of the NVT run at 1250 K.
The pair correlation function,  the distribution of the coordination numbers and the 
 bond angle distribution function of liquid Ge from NN and DFT simulations are compared in Fig. \ref{GR_GE_LIQ}. 

\begin{figure*}[htbp!]
    \centering
    \includegraphics[scale= 0.6]{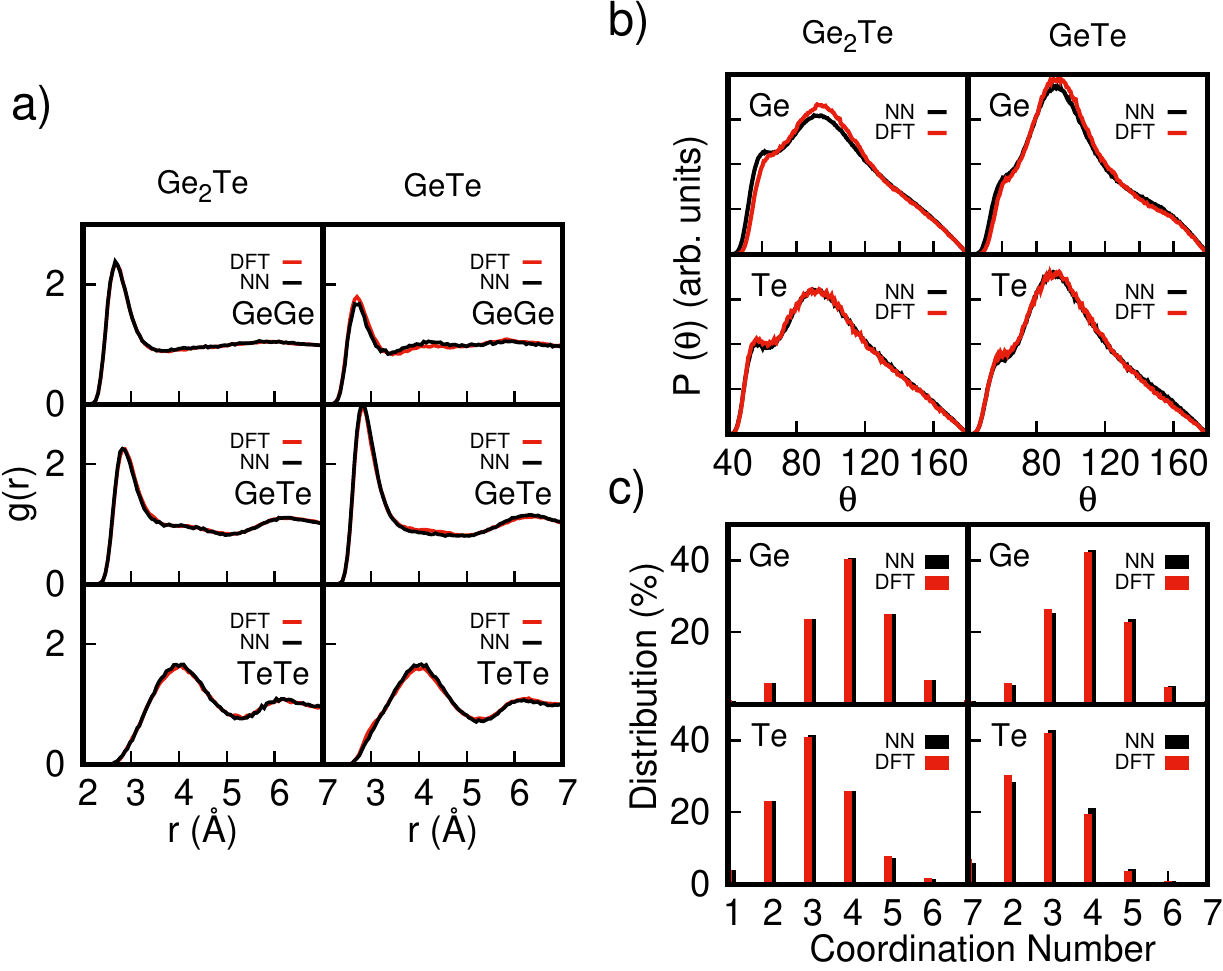}
    \caption{Structural properties of  liquid Ge$_2$Te at 1200 K (left panels) and liquid GeTe at 1150 K (right panels) from DFT and NN simulations. a) Partial pair correlation functions. b) Bond angle distribution.  c) Distribution of coordination numbers. }
    \label{GR_GETE_LIQ}
\end{figure*}

The structural properties of liquid GeTe  (4096-atom model) and of liquid  Ge$_2$Te (2400-atom model) from NN-MD simulations were compared with  DFT-MD results for 300-atom models  for both compositions. For GeTe, we used the experimental equilibrium density of the liquid phase at 1150 K of 0.03294 atom/\r{A}$^3$ \cite{mazzaGeTeliquid}. Since no experimental data are available for the density of liquid Ge$_2$Te, and in the perspective to study the amorphous phase, we generated an amorphous model with DFT molecular dynamics by quenching from 1200 K to 300 K in 100 ps in the NPT ensemble which yielded an equilibrium density at 300 K of 0.0355 atom/\r{A}$^3$. To this aim, we used the Grimme (D3) semiempirical potential \cite{D3} for van der Waals (vdW) interactions which is needed to avoid the coalescence of nanovoids in the liquid phase as discussed in Ref. \cite{sosso2012breakdown}. Liquid Ge$_2$Te was then simulated without vdW corrections at the same density of the amorphous phase of 0.0355 atoms/\r{A}$^3$. 
The liquid models were equilibrated first at 2000 K for 10 ps and then  for 20 ps at 1150 K for GeTe and at 1200 K for Ge$_2$Te, which are well above the melting temperature (990 K for 
the melting of GeTe \cite{liquidGeTe} and for the incongruent melting of Ge$_2$Te, the liquidus temperature of Ge$_2$Te is instead 1150 K \cite{GeTePhaseDiagram}). Structural properties were calculated in the last 10 ps.
The pair correlation functions, the distribution of the coordination numbers and the bond angle distribution of liquid Ge$_2$Te and GeTe  from NN and DFT simulations are compared in Fig. \ref{GR_GETE_LIQ}.
The bonding cutoff of 3 \r{A} for Ge-Ge, 3.22 \r{A} for Ge-Te and 3 \r{A} for Te-Te have been used.
The average partial coordination numbers are shown in Table \ref{coord_liquid}. 

Overall, the NN potential reproduces very well the structural properties of the liquid phase for all the three compositions, i.e. pure Ge, GeTe and Ge$_2$Te.

\begin{table}[htbp!]
    \centering
    \begin{tabular}{lcccc}
        \hline
    \hline
      && Ge & Te & Total \\
 \hline
 Ge$_2$Te   &Ge&2.54 (2.53)&1.50 (1.51)&4.04 (4.04)\\
            &Te&3.05 (3.07)&0.07 (0.07)&3.12 (3.14)\\
     \hline
 GeTe       &Ge& 1.25 (1.31) & 2.73 (2.63)&3.98 (3.94)\\
            &Te&2.73 (2.63)&0.16 (0.19)&2.89 (2.82)\\
 \hline
     \hline
    \end{tabular}
    \caption{Average partial coordination numbers in liquid Ge$_2$Te at 1200 K and liquid GeTe at 1150 K. DFT data are reported in parenthesis.} 
    \label{coord_liquid}
\end{table}

Regarding the dynamical properties, we computed the diffusion coefficient ($D$) in liquid GeTe and Ge$_2$Te from the mean square displacement (MSD) and the Einstein relation $MSD$=$6 Dt$ in 30 ps simulations as given in Table \ref{diff_NN}.

\begin{table}[htbp!]
    \centering
    \begin{tabular}{lcccc}
    \hline
    \hline
    && \multicolumn{2}{c}{ D (10$^{-5}$ cm$^2$/s)} \\
      && {Ge} & {Te} \\
\hline
 Ge$_2$Te   &1100 K& 9.5 (8.7) & 6.1 (5.9) \\
 GeTe       &1120 K& 8.0 (6.3) &  6.0   (5.0)\\
 \hline
 \hline
    \end{tabular}
    \caption{Diffusion coefficients in liquid Ge$_2$Te at 1100 K and liquid GeTe at 1120 K from NN and DFT (in parenthesis) simulations.} 
    \label{diff_NN}
\end{table}
Both in Ge$_2$Te and GeTe, the NN potential slightly overestimates the diffusion coefficients by about 10\%.
Part of this discrepancy may come from the use of  different simulation cells in the NN and DFT calculations. 
In fact, the following scaling of $D$ with the size $L$ of the cell is expected \cite{kremer}

\begin{equation}
D(L)= D_{\infty}-\frac{2.387 k_B T}{6 \pi \eta L},
\label{scaling}
\end{equation}

where $\eta$ is the viscosity. For $\eta$ of about 1 mPa$\cdot$s
(see the NN results at 1000 K of Ref. \cite{sosso2012breakdown}) the correction to $D$ for liquid GeTe at 1120 K is about
0.9 10$^{-5}$ cm$^2$/s for a 300-atom cell ($L$=20.88 \AA)
and 0.375 10$^{-5}$ cm$^2$/s for a 4096-atom cell ($L$=49.91 \AA).

\begin{figure}[htbp!]
    \centering
    \centerline{\includegraphics[width=0.48\textwidth]{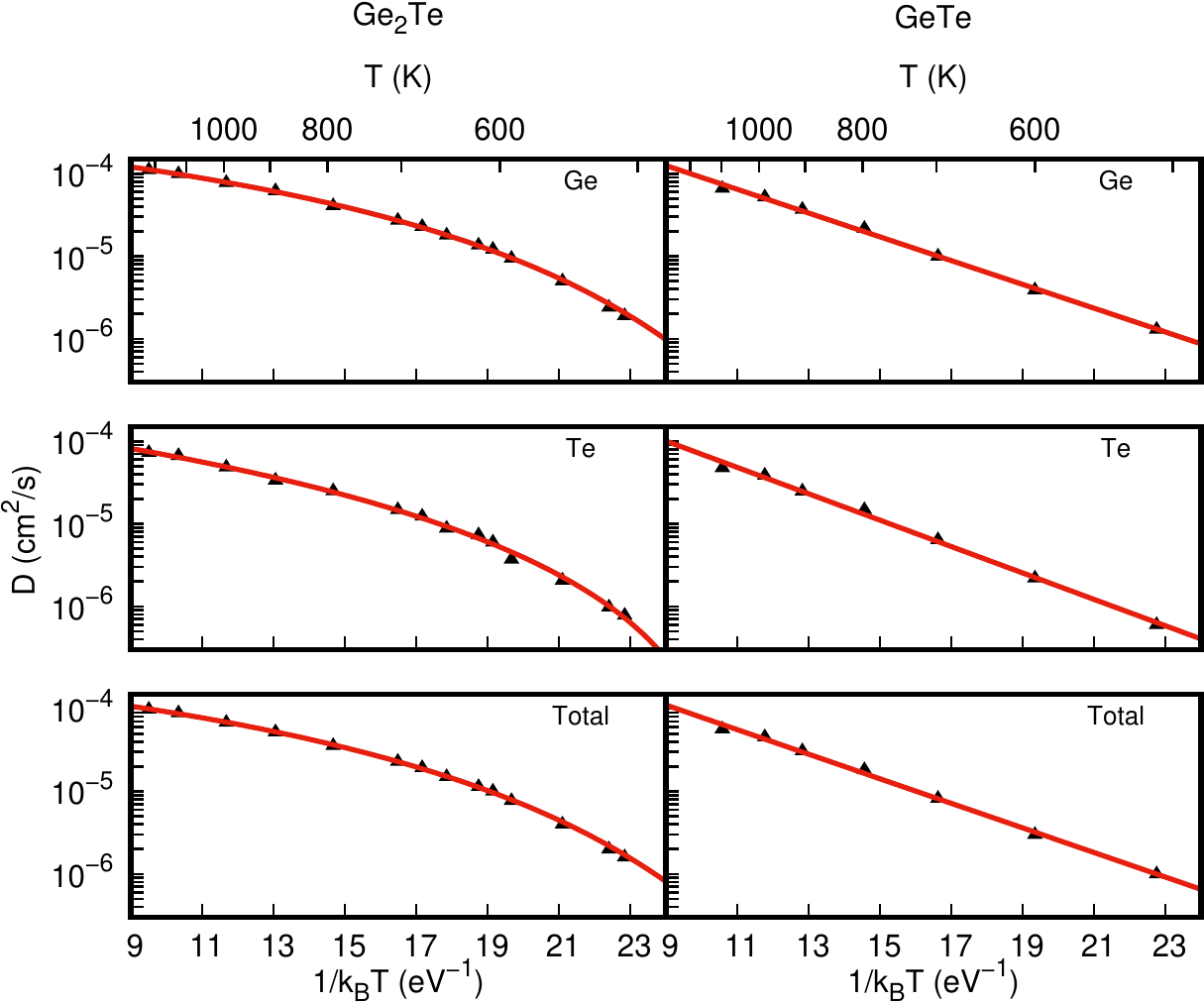}}
    \caption{Diffusion coefficient ($D$) as a function of temperature from NN simulations of GeTe and Ge$_2$Te for  (upper panels) Ge atoms only, (central panels) Te atoms only and (bottom panels) the total values averaged over all atoms. The continuous lines are Cohen-Grest fit (see text) of the data for Ge$_2$Te and a simple Arrhenius fit for GeTe.}
    \label{cgdiff}
\end{figure}

The diffusion coefficient (total and resolved per species) has then been computed on a wide temperature range in the liquid and supercooled liquid phases from NN simulations, 50 ps long at each temperature, as shown in Fig. \ref{cgdiff}. 
The diffusion coefficient of Ge$_2$Te is fitted by a  Cohen-Grest  function \cite{CG} $\log(D(T))=A-{2B}/{(T-T_0 + [(T-T_0)^2 + 4CT ]^{1/2})}$  which is suitable for the supercooled phase of fragile liquids. According to Angell~\cite{Angell}, a supercooled liquid is classified as fragile if the viscosity $\eta$ remains
very low down to temperatures close to the glass transition temperature T$_g$, where a steep,
super-Arrhenius behavior is observed up to the high value of $\eta$ expected at T$_g$. A similar super-Arrhenius behavior is expected for the diffusion coefficient $D$ given its inverse proportionality with the viscosity $\eta$ according to the Stokes-Einstein relation.
On the contrary the viscosity of an ideal strong liquid follows an Arrhenius behavior from the melting down to T$_g$. 
The  fitting parameters for Ge$_2$Te amount to $A$= -3.52 (-3.49), $B$= 435 K (413 K), $C$= 0.36 K (0.52 K) and $T_0$= 315 K (321 K)  with the values in parenthesis referring to the diffusion coefficient of Ge atoms only. 
For GeTe, the diffusion coefficient follows an Arrhenius behaviour D = D$_0$e$^{-E_a/(k_BT)}$ down to 500 K as shown in Fig. \ref{cgdiff} and in agreement with previous NN results in Ref. \cite{sosso2012breakdown}, due to the breakdown of the Stokes-Einstein relation at low temperatures. {The fitting parameters for GeTe are E$_a$ = 0.34 eV and D$_0$ = 2.47 $\cdot$ 10$^{-3}$ cm$^2$/s.} The super-Arrhenius behaviour in $D$ for GeTe would appear at lower temperatures, below 500 K, which implies a higher fragility of GeTe with respect to Ge$_2$Te  considering that T$_g$ is expected to be higher for Ge$_2$Te than for GeTe. The reported glass transition temperature for pure germanium of about 750 K \cite{Imabayashi} is indeed higher than the value of 423 K reported  for GeTe \cite{chenTgGeTe}. 
For the sake of comparison the diffusion coefficient of pure Ge as a function of temperature is given in Fig. S2 in the Supplementary Information.

\subsubsection{The amorphous phase}
\begin{figure*}[htbp!]
    \centering
    \centerline{\includegraphics[scale= 0.8]{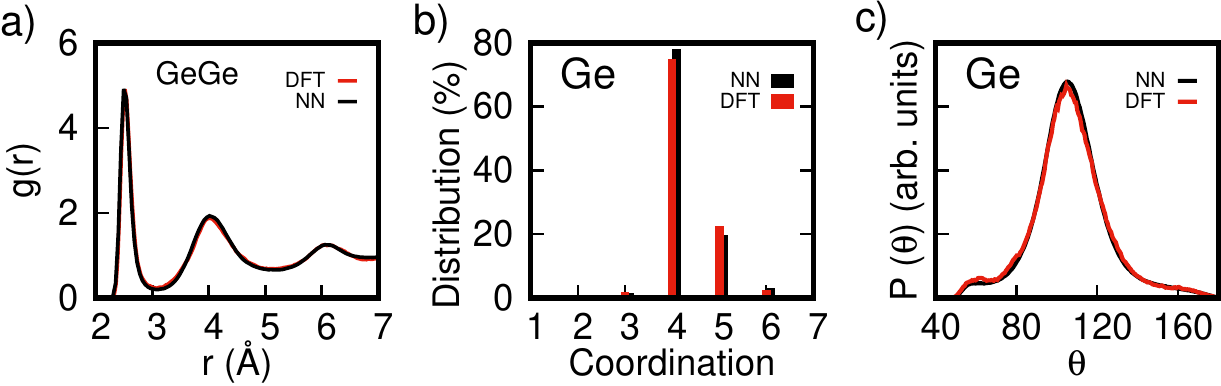}}
    \caption{Structural properties of amorphous Ge at 300 K from NN and DFT simulations. a) Pair correlation function. b) Distribution of coordination numbers by assuming a bond cutoff of 3 \r{A}. 
    c) Bond angle distribution function.}
    \label{GR_GE}
\end{figure*}

An amorphous model of Ge  was generated by cooling the liquid model from 1250 K to 300 K in 140 ps in both the DFT and NN simulations at the experimental density of a-Ge of 0.0438 atoms/\AA$^3$ \cite{baribeau}. Structural properties were averaged over the last 10 ps of an equilibration run at 300 K lasting 20 ps. NN and DFT results for the pair correlation function, the distribution of the coordination numbers and the bond angle distribution function are compared in Fig.\ref{GR_GE}. 

\begin{figure*}[htbp!]
    \centering
    \includegraphics[width=1\textwidth]{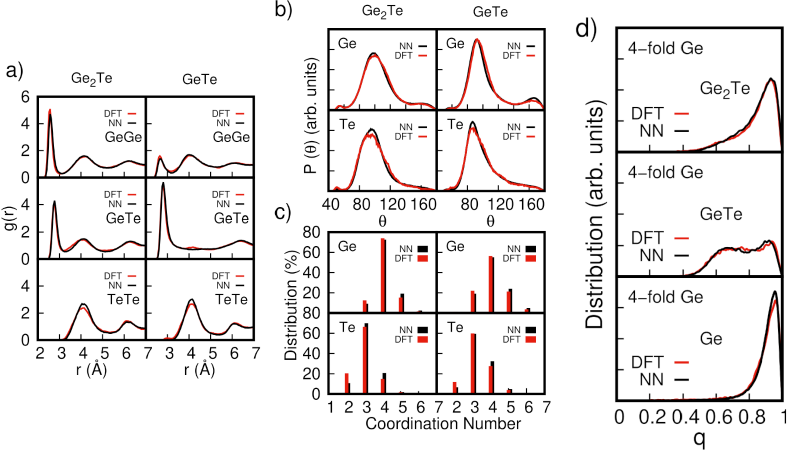}
    \caption{Structural properties of  amorphous Ge$_2$Te (left panels)  and GeTe (right panels) at 300 K from DFT and NN simulations. a) Partial pair distribution functions. b) Bond angle distribution functions. c) Distribution of the coordination numbers. d) Distribution of the $q$ order parameter for tetrahedricity, also for pure Ge.}
    \label{GR_GETE}
\end{figure*}

Amorphous models of Ge$_2$Te (a-Ge$_2$Te) and GeTe (a-GeTe) were obtained by quenching the liquid to 300 K in 100 ps starting from 1200 K  for Ge$_2$Te and from 1150 K for GeTe in both NN and DFT simulations. The pair correlation functions, the distribution of the coordination numbers, and the bond angle distribution functions are compared in Fig.\ref{GR_GETE}, while  the average partial coordination numbers are given in Table \ref{COORD_AM}. The bonding cutoffs are the same used for the liquid as given above. 

The agreement between the NN and DFT results is overall very good for Ge, GeTe and Ge$_2$Te. We just note a slight misfit in the number of 2-fold Te atoms in a-Ge$_2$Te and in the position and intensity of the peak of the Ge-Ge correlation function in a-GeTe, possibly due in part to the variability  from model to model in small 300-atom cells as discussed in a previous work on stoichiometric GeTe \cite{SossoNN}.  

\begin{table}[htbp!]
    \centering
    \begin{tabular}{lcccc}
    \hline
    \hline
      && Ge & Te & Total \\
\hline
 Ge$_2$Te   &Ge&2.59 (2.58)&1.54 (1.46)&4.13 (4.04)\\
            &Te&3.12 (2.96)&0.00 (0.00)&3.12 (2.96)\\
     \hline
 GeTe       &Ge&0.79 (0.87)&3.34 (3.18)&4.13 (4.04)\\
            &Te&3.34 (3.18)&0.00 (0.04)&3.34 (3.21)\\
   \hline
   \hline
    \end{tabular}
    \caption{Average partial coordination numbers in amorphous Ge$_2$Te and GeTe at 300 K from NN and DFT (in parenthesis) simulations.} 
    \label{COORD_AM}
\end{table}

Turning now to a comparison of the structural properties of Ge$_2$Te and GeTe, we notice that by increasing the fraction of Ge from GeTe to Ge$_2$Te, we observe also an increase in the fraction of Ge-Ge bonds that favor a tetrahedral environment of Ge \cite{deringer2014bonding}. This feature shows up as an increased fraction of 4-fold coordinated Ge in the distribution of the coordination numbers (see Fig. \ref{GR_GETE}) and in the shift to higher angles of the peak in the bond angle distribution. In general, we expect  a coexistence of tetrahedral, pyramidal (3-fold coordinated) and defective octahedral coordination (octahedral angles but coordination lower than six)  for Ge atoms. The fraction of Ge atoms in tetrahedral configurations can be quantified by using the $q$-parameter for tetrahedricity introduced in Ref. \cite{qparam} and defined by 
$  q_j=1-\frac{3}{8}\Sigma_{i<k}(\frac{1}{3}+cos(\theta_{ijk}))^2$,
where the sum runs over the couples of atoms bonded to a central atom $j$ and forming a bonding angle $\theta_{ijk}$. 
The order
parameter evaluates to $q$=1 for the ideal tetrahedral geometry, to $q$=0 for
the 6-fold coordinated octahedral site, to $q$=5/8 for a 4-fold coordinated
defective octahedral site, and $q$=7/8 for a pyramidal geometry.
The distribution of the $q$-parameter for 4-coordinated Ge atoms is shown in Fig. \ref{GR_GETE}d for amorphous Ge$_2$Te, GeTe and Ge at 300 K. The $q$-parameter features two peaks in a-GeTe due to the coexistence of tetrahedral 
and defective octahedral configurations \cite{mazza}. As discussed in Ref. \cite{spreafico}, the integration of the $q$-parameter in the range 0.8-1.0 gives a good measure of the fraction of tetrahedral Ge which is 24\% in GeTe, similar to previous DFT works \cite{akola2007structural,mazza}. On the other hand,  a-Ge$_2$Te  features a single  peak centered at the position corresponding to the tetrahedral configuration (as in a-Ge). The peak is, however,  very broad which suggests that also in Ge$_2$Te a percentage of Ge atoms are in  defective octahedral configurations. By integrating the $q$-distribution for values greater than 0.8, one obtains a percentage of tetrahedral Ge of approximately 52\% in a-Ge$_2$Te , which is in between the values of 24\% and 68\% for a-GeTe and a-Ge.
Analogously to the liquid phase, the agreement between the NN and DFT results for the amorphous phase is overall very good for Ge, GeTe and Ge$_2$Te.

\subsubsection{The crystalline phase}
We calculated the energy of crystalline  cubic Ge and of trigonal ($\alpha$-phase) GeTe at different volumes and we fitted the energy as a function of volume by using the Birch-Murnaghan equation of state as shown in Fig. S3 in the Supplementary Information. The resulting fitting parameters from NN and DFT calculations are compared in Table \ref{EOS_cubic_Ge} while
the equilibrium structural parameters of $\alpha$-GeTe (space group $R3m$)\cite{GeTe} are given in Table \ref{trigonal_GeTe}.
Crystalline  $\alpha$-GeTe, with two atoms per unit cell, can be viewed as a distorted rocksalt geometry with an elongation of the cube diagonal along the [111] direction and an off-center displacement of the inner Te atom along the [111] direction giving rise to a 3+3 coordination of Ge with three short and stronger bonds  and three long and weaker  bonds (see Table \ref{trigonal_GeTe}).  In the conventional hexagonal unit cell of the trigonal phase, the structure can be also seen as an arrangement of GeTe bilayers along the $c$ direction with shorter intrabilayer bonds and longer interbilayers bonds. We mention that the trigonal ferroelectric phase transforms  into the cubic paraelectric phase ($\beta$-phase, space group Fm$\bar{3}$m) above the Curie temperature of 705 K \cite{BGeTe}. In the cubic phase, the alternation of long and short bonds  survives in a disordered manner along all equivalent $<$111$>$ directions as revealed by extended x-ray absorption fine structure  (EXAFS), x-ray total diffraction measurements \cite{TransGeTe,matsunagabeta} and MD simulations \cite{mazzarellobeta}.
However, more recent molecular dynamics simulations \cite{fahy} suggest that the order-disorder character of the phase transition is weaker than as inferred from EXAFS data.
The $\beta$-phase is the structure the a-GeTe crystallizes into at the operation conditions of the memory devices.

\begin{table}[htbp]
\centering
\begin{tabular}{ccccc}
\hline
\hline
&& cubic Ge \\
\hline
\hline
Method & E$_0$ (eV/atom)& V$_0$ (\AA$^3$/atom)& B (GPa) & B$^\prime$ \\
\hline
NN&   -106.972&24.22&56.99&4.56\\
DFT&  -106.972&24.22&57.58&4.64\\
\hline
\hline
&& $\alpha$-GeTe \\
\hline
\hline
Method & E$_0$ (eV/atom)& V$_0$ (\AA$^3$/atom)& B (GPa) & B$^\prime$ \\
\hline
NN&   -164.671&28.33&29.27&8.42\\
DFT & -164.671&28.34&28.88&8.55\\
\hline
\hline
\end{tabular}
\caption{Fitting parameters of the Birch–Murnaghan equation of state of cubic Ge and trigonal GeTe. V$_0$ is the equilibrium volume, E$_0$ the equilibrium energy, B$_0$ the bulk modulus and B$^\prime$ its derivative respect to pressure.}
\label{EOS_cubic_Ge}
\end{table}

\begin{table}[htbp]
\centering
\begin{tabular}{cccccccc}
\hline
\hline
Method & $a$ (\r{A}) & $\alpha$ & $V$ (\r{A}$^3$) & $x$ & $d_{short}$ (\r{A})& $d_{long}$ (\r{A})\\
\hline
NN&   4.42 & 56.95\degree & 57.00 & 0.2340 & 2.86 & 3.29 \\
DFT & 4.33 & 58.14\degree & 55.00 & 0.2358 & 2.85 & 3.21 \\
\hline
\hline
\end{tabular}
\caption{Equilibrium structural parameters of the $\alpha$-phase of GeTe in the rhombohedral setting from NN and DFT calculations. $a$ is the lattice parameter, $\alpha$ is the angle of the trigonal cell, $V$ is the unit cell volume (two atoms), $x$ assigns the position of Ge atom at $(x,x,x)$ and Te atom at $(-x,-x,-x)$ in crystallographic units. $d_{short}$ and $d_{long}$ are the lengths of the short and long Ge-Te bonds.}
\label{trigonal_GeTe}
\end{table}

In summary, although the RMSE for the energies and forces are not very small (4.4 meV/atom and 105 meV/\r{A}), albeit similar to other NN potentials in literature for disordered multi-component materials, the validation of the potential over the properties of liquid, amorphous and crystalline phases is excellent. Overall, we judge that our potential is sufficiently accurate to address the study of the crystallization process as discussed in the next section.

\subsection{Simulation of the crystallization process}

We exploited the NN potential discussed above to perform simulations of the crystallization process in both GeTe and Ge$_2$Te. 
We first simulated stoichiometric GeTe to compare the results with previous NN simulations in Ref. \cite{SossoCryst}.

\subsubsection{GeTe}
\begin{figure*}[htbp!]
    \centering
    \includegraphics[width=1\textwidth]{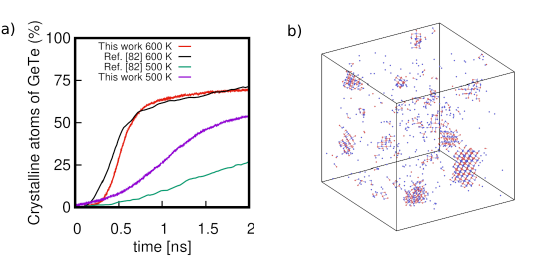}
    \caption{a) Fraction of crystalline atoms (defined by Q$^{dot}_{4}$)  as a function of time in a 32768-atom model of GeTe in simulations at 500 and 600 K, compared with previous results (4096 atoms) from Ref. \cite{SossoCryst}. b) Snapshot of the crystalline atoms in the 32768-atom model of GeTe after 100 ps at 600 K.}
    \label{crystal_gete}
\end{figure*}

We generated a 32768-atom model of a-GeTe at the experimental amorphous density of 0.0333 atom/\r{A}$^3$ \cite{ghezzi}. The model was  equilibrated first at 2000 K for 10 ps and  at 1150 K for 40 ps, and it was then cooled to either 600 or 500 K in 60 ps. Finally, we performed a NVT simulation at the two target temperatures. 
 The number of crystalline atoms as a function of time 
at 600 and 500 K are compared in Fig. \ref{crystal_gete}a with the results from previous NN simulations  \cite{SossoCryst} at the same temperatures and with 4096 atoms.
At 600 K,  our data are very similar to previous results \cite{SossoCryst} with a fraction of crystallized atoms of almost 75 \%   after 2 ns. 
 Some small crystalline nuclei of cubic GeTe form and begin to grow after the first 100 ps, as shown in the snapshot of Fig. \ref{crystal_gete}b from the simulation at 600 K.
At 500 K,  the number of crystalline atoms increases faster with time in our simulation than  in the previous one from Ref. \cite{SossoCryst}, because our system is eight times larger than that of Ref. \cite{SossoCryst} and then the nucleation time  (time needed for the formation of an overcritical  crystalline nucleus) is shorter. 
We can thus conclude that our potential reproduces the crystallization kinetics in a manner very similar to the NN potential of Ref. \cite{SossoCryst} which was previously used to address several details of the crystallization process in GeTe \cite{SossoCryst,gabardi2017atomistic,Gabardi2019,perego,Sosso2015,Acharya}.

Regarding the mechanism of crystal nucleation and growth, we mention that in a very recent work \cite{yarema} it has been proposed that crystallization is triggered by a pre-ordering of the face-centered-cubic (fcc) sublattice of the Te atoms that precedes the ordering of the square rings of Ge-Te bonds.  This picture was inferred from the analysis of the temperature/time dependence of the Te-Te coordination numbers (first neighbors on the anionic fcc sublattice) obtained from the fitting of extended x-ray absorption fine structure (EXAFS) spectrum at the Te K-edge. To check this picture, we computed the   Q$^{dot}_{6}$ order parameter for the Te-Te coordination involving only Te atoms and Te-Te contacts at distances up to  4.4 \AA\ which is slightly above  the nearest neighbor distance on the Te fcc sublattice. The Q$^{dot}_{6}$ for Te atoms only would thus detect the ordering of the Te fcc sublattice. On the contrary, the  Q$^{dot}_{4}$ parameters involving pairs of atoms up to 3.6 \r{A} measures the ordering of the simple cubic lattice made of the two fcc sublattices of Ge and Te, i.e. it detects the ordering of the square rings made of Ge-Te bonds.  A pre-ordering of the Te fcc sublattice 
should then result into an increase in the number of crystalline Te atoms assigned by the Q$^{dot}_{6}$ order parameter before cristallinity would be detected by the Q$^{dot}_{4}$ order parameter which is tuned on the ordering of Ge-Te bonds.
The evolution in time of the crystalline Te atoms assigned by the Q$^{dot}_{6}$  and Q$^{dot}_{4}$
parameters is shown in Fig. S4 in the Supplementary Information. No delay is observed in the onset of the
growth of Q$^{dot}_{4}$ with respect to Q$^{dot}_{6}$  which means that there is no pre-ordering of the Te fcc sublattice.  We remark, however, that the experimental analysis in Ref. \cite{yarema} was performed at 440 K on a transformation occurring on the time scale of hours, while our analysis addresses the nucleation and growth occurring on the time scale of hundreds of ps at 500-600 K. Therefore, we can not  exclude that pre-ordering of the Te fcc sublattice could take place at the experimental conditions mentioned above, albeit according to our results this process does not seem to occur at the operation conditions of PCMs. For our purpose, we considered this validation sufficient to move to the study of the crystallization in Ge$_2$Te which is discussed hereafter.

\subsubsection{Ge$_2$Te} 
As mentioned in the Introduction, crystallization of the Ge$_{63}$Te$_{37}$ alloy deposited by magnetron sputtering was studied by several means in Ref. \cite{carria}. There, it was found that the crystallization process consists of three 
separate steps: first, Ge segregates by forming a-Ge regions followed by the crystallization of GeTe and later of  pure Ge. These evidences resulted from time resolved measurements of optical reflectivity, x-ray diffraction and Raman spectra during an annealing ramp (0.1 K/s)  \cite{carria}. The transformation starts at about 600 K,  GeTe crystallization was observed at about 630 K, while crystalline Ge appears at an even  later time  at about 650 K. 
The crystallized sample was also re-amorphized by laser melting
and the resulting amorphous sample crystallizes at the same temperature of the asdep one. In another work \cite{raoux2}, a longer crystallization time of about one order of magnitude was reported instead for asdep Ge$_{62}$Te$_{38}$ sample than for the melt quenched one. This difference was ascribed to the presence of an amorphous-crystal interface in the laser melted sample which favors crystal growth without the need of crystal nucleation \cite{raoux2}.
A previous Raman study confirms a multistep crystallization process for Ge$_{76}$Te$_{24}$ and  Ge$_{62}$Te$_{38}$ \cite{gourvest}. GeTe crystallites were detected at 470 K in Ge$_{62}$Te$_{38}$ whereas pure Ge crystallizes at a later time at 650 K \cite{gourvest}.
A similar behaviour  was reported from  time-resolved diffraction measurements in Ref. \cite{raoux} for Ge$_{62}$Te$_{38}$ and in Ref. \cite{navarro2013SSE} for Ge$_{62}$Sb$_{38}$ and Ge$_{69}$Sb$_{31}$. However, the possibility of the formation of large a-Ge region due to Ge segregation  prior to crystal nucleation of GeTe, was not discussed explicitly in Refs. \cite{gourvest,raoux,navarro2013SSE}.
A simultaneous crystallization of GeTe and Ge was reported instead at high temperature of 620 K for
Ge$_{70}$Te$_{30}$ in Ref.
\cite{raoux} under an annealing ramp of 1 K/s.
The reason behind the discrepancies among different reports is unclear. Different degree of homogeneity of the alloy under different preparation conditions or different levels of surface oxidation may affect the crystallization temperature. 
Indeed, oxidation was shown to sizeably  lower T$_x$ in GeTe \cite{Berthier2017}  and in Ge-rich GST alloys \cite{agatioxide}. 

On these premises, we started our analysis by first simulating Ge$_2$Te at 600 K  in
a 19200-atom model  at the theoretical amorphous density of 0.0355 atom/\r{A}$^3$ (see Sec. III.A.1). 
By assuming that Ge would segregate as a-Ge at the experimental density of 0.0438 atom/\r{A}$^3$, a phase separation of Ge$_2$Te at constant volume and at the average density of 0.0355 atom/\r{A}$^3$ would lead to GeTe regions at the density of 
0.0326 atom/\r{A}$^3$ which is sufficiently close to the experimental density of a-GeTe
of 0.0333 atom/\r{A}$^3$ \cite{ghezzi}. Therefore all simulations were performed at constant volume which is also closer to the operation conditions of PCMs.
The 19200-atom model was equilibrated first at 2000 K for 10 ps and at 1200 K for 40 ps and then quenched  to 600 K in 60 ps. At this temperature we performed a NVT simulation lasting 14 ns.

\begin{figure*}[ht!]
    \centering
      \centerline{\includegraphics[scale= 1.2]{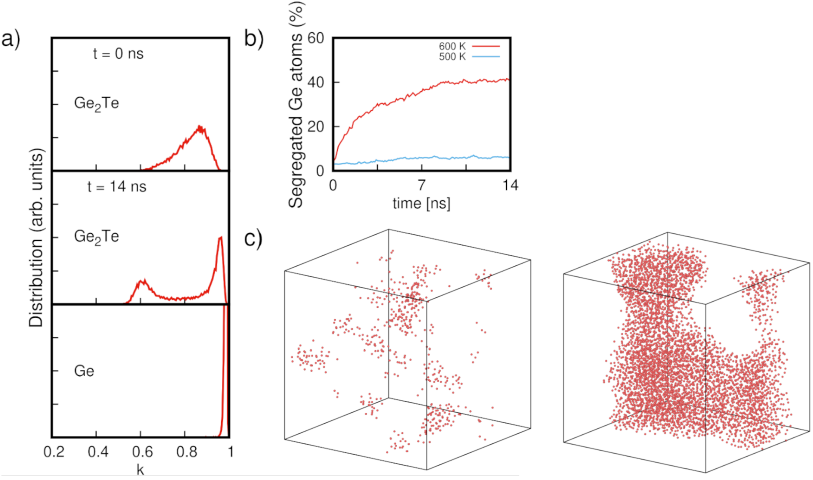}}
    \caption{a) Distribution of the SOAP $k_j$ at the beginning (top, t = 0 ns ) and after 14 ns (center) in the simulation of Ge$_2$Te (19200 atoms) at 600 K  compared to the distribution of an independent model of a-Ge at the same temperature (bottom). b) Number of segregated Ge atoms ($k_j$ $>$ 0.92) as a function of time in the  19200-atom  model of Ge$_2$Te at 600 K (red) and 500 K (blue). c) Snapshots of  Ge$_2$Te at the beginning (left) and after 14 ns  (right) at 600 K. Only atoms with an environment close to that of a-Ge are shown (SOAP $k_j$ $>$ 0.92).}
    \label{segreg_snap}
\end{figure*}

At 600 K, Ge atoms immediately start to segregate from Ge$_2$Te. Phase separation occurs because the sum of the free energies of  GeTe and Ge is lower than that of Ge$_2$Te in the amorphous of supercooled liquid phases. Indeed, we verified that the reaction energy of the transformation Ge$_2$Te $\rightarrow$ GeTe + Ge at 600 K is $\Delta$E = 7 $\pm$ 2 meV/atom, as obtained from 6 independent simulations of supercooled liquid models of these compositions at 600 K. Although we did not attempt to estimate the reaction entropy, the results above confirm that Ge$_2$Te is unstable against phase separation.
To quantify the fraction of segregated Ge atoms, we computed the SOAP similarity kernel $k_j$ for a-Ge (SOAP $k_j$) as described in Sec. II. SOAP $k_j$ for a Ge atom is close to one for configurations similar to the average local configuration in a-Ge and it decreases by
increasing the dissimilarity with a-Ge. 
The distribution of $k_j$ at the beginning (t = 0 ns) and at the end (t = 14 ns) of the simulation is compared in Fig. \ref{segreg_snap}a  to the distribution $k_j$ of a-Ge at the same temperature. 
We could qualitatively assess that a Ge atom is segregated  when its SOAP $k_j$ is higher than 0.92, as the distribution of $k_j$ in a-Ge ranges from 0.92 to 1.
In the simulation of Ge$_2$Te,  the $k_j$ distribution initially features a single broad peak at around 0.9, while after 14 ns two peaks are present at 0.6 and  at 0.95 which highlights the occurrence of a phase separation.
As shown in Fig. \ref{segreg_snap}b
 the  number of segregated Ge atoms increases initially very fast and then it levels off at about 40 $\%$ after 10 ns.
 Complete phase separation into Ge and GeTe  would correspond to a fraction of 50 $\%$ of segregated Ge atoms.
The segregation of Ge is very clear in the snapshots reported in Fig. \ref{segreg_snap}c
where we show only Ge atoms in an environment close to a-Ge 
($k_j$ $>$ 0.92) at the beginning of the simulation and after 14 ns.

Segregation of  Ge  leads to the formation of Ge-rich regions with Ge$_{88}$Te$_{12}$ composition and average  density of 0.042 atom/\r{A}$^3$, and of Ge-poor regions with  Ge$_{53}$Te$_{47}$ composition  and average density of 0.0327 atom/\r{A}$^3$ which is very close to the experimental density of the amorphous phase of stoichiometric GeTe (0.0333  atom/\r{A}$^3$). 
 The composition of the Ge-poor region was computing by averaging over about 4000 atoms in a region at least 10 \r{A} far from the region of segregated Ge (see Fig. \ref{segreg_snap}). Similarly, the composition of the Ge-rich region was computed by averaging over about 3000 atoms
 in the lower part of the a-Ge-like region (see Fig. \ref{segreg_snap}).

 We also performed a second  simulation starting from an independent 
19200-atom model at 600 K with very similar results. The snapshot of the
segregated atoms at the beginning and the end (14 ns) of this second simulation is shown in Fig. S5 in the Supplementary Information.
In the process of formation of the a-Ge regions, Ge atoms is the most diffusing specie as one would envisage from the diffusion coefficients in Fig. \ref{cgdiff}. The overall average MSD over 10 ns  for each species is indeed very close to the value given by the Einstein relation. At 600 K the square root of the average MSD after 10 ns is about 8.5 nm for Ge atoms and about 5.9 nm for Te atoms. We remark that the edge of the simulation cell is 8.1 nm.
 
\begin{figure*}[htbp!]
    \centering
    \centerline{\includegraphics[width=1\textwidth]{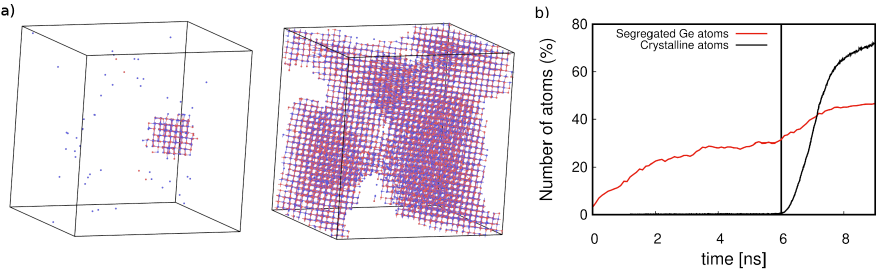}}
    \caption{ a) Crystalline atoms in the segregated large model (30000 atoms) of Ge$_2$Te at t = 6 ns  and at t= 10 ns at 600 K. b) Fraction of crystalline atoms (defined by Q$^{dot}_{4}$, black line)  and number of segregated Ge atoms ($k_j >$ 0.92, red line) as a function of time.}
    \label{crystal_ge2te_big}
\end{figure*}

Although most of Ge atoms in excess is segregated, we do not see the formation of crystalline nuclei in the Ge$_{53}$Te$_{47}$ region in the simulations lasting 14 ns. We can conclude that at this temperature the nucleation time in a 19200-atom cell is longer than 14 ns. 
Raising the temperature from 600 K to 630 K for a 4 ns does not lead to further segregation of Ge nor to crystal nucleation. 
In the attempt to reduce the crystal nucleation time, we simulated a larger 30000-atom model prepared with the same protocol of the smaller ones. 
We chose an orthorhombic cell with edges a=b= 101.82 \r{A} and c= 81.45 \r{A}. Phase separation occurs in a similar manner as shown in Fig. S5 in the Supplementary Information.
In the larger model, however, an overcritical crystalline nucleus of GeTe appears at 600 K after 6 ns and it grows until the crystal percolates through the simulation box as shown in Fig.  \ref{crystal_ge2te_big}.

Once formed, the overcritical crystalline nucleus grows,  but at a lower rate compared to stoichiometric GeTe at the same temperature. 
 At 600 K, we calculated the crystal growth velocity $v_g$=$dR/dt$, where $R$ is the radius of the crystalline nucleus given in turn by
 $R\left(t\right)=\left({3N(t)}/4\pi\rho\right)^{\frac{1}{3}}$, where $N$ is the number of atoms in the  nucleus and $\rho$ is the density of the crystalline phase (0.0351 atoms/\r{A}$^3$). This assumption is valid only in the early stage of crystallization when the nuclei do not interact with each other  or with their periodic image. As an example, 
 the evolution of $R(t)$ as a function of time is shown  Fig. S6 in the Supplementary Information  for a single crystalline nucleus in GeTe and Ge$_2$Te (30000-atom cell) at 600 K.  In GeTe, we computed $v_g$ by
 averaging over three/(four) crystalline nuclei at 600 K/(500 K).  
 The resulting v$_g$ are reported in Table \ref{velocities}.
 The crystal growth velocity is lower in Ge$_2$Te because a further segregation of Ge must take place during the crystal growth in region with average composition of Ge$_{53}$Te$_{47}$.
 The average composition of the crystallized region is in fact Ge$_{50}$Te$_{50}$.
 
\begin{table}[htbp!]
\centering
\begin{tabular}{lccc}
\hline
\hline
& \multicolumn{2}{c}{$v_{\rm g}$ (m/s)} \\

 & 500 K & 600 K\\
\hline
GeTe (homogeneous) &  1.2&  4.8  \\
Ge$_2$Te (phase separated, large)&  & 3.1 \\
Ge$_2$Te (phase separated, small)& 0.8  &   \\
Ge$_2$Te (homogeneous, small)& 0.2  &  \\
Ge$_{53}$Te$_{47}$ (homogeneous)& 0.8  &  \\
\hline
\hline
\end{tabular}
\caption{Crystal growth velocities v$_g$ of overcritical nuclei at 500 K or 600 K in the different models. Homogeneous GeTe; 
phase separated Ge$_2$Te (large 30000-atom cell); Ge$_2$Te (small 19200-atom cell) phase separated at 600 K, quenched at 500 K to promote crystal nucleation; homogeneous Ge$_2$Te before Ge segregation (small 19200-atom cell);  homogeneous Ge$_{53}$Te$_{47}$.}
\label{velocities}
\end{table}
  
We also attempted  to reduce the crystal nucleation time  in the 19200-atom smaller segregated model by quenching  to 500 K. Indeed, at this temperature,  we see the formation of three crystalline nuclei after some hundreds of ps in the Ge$_{53}$Te$_{47}$ region, as shown in Fig. S7 in the Supplementary Information. This occurs at positions where the local Ge-Ge coordination is lower than the average in a-GeTe (0.79). The strong shortening of the nucleation time  at 500 K, once the system was mostly phase separated at 600 K, might be due to the smaller size of the critical nucleus at lower temperatures. Since the local Ge content must be lower than the average composition in the Ge-poor region (at incomplete phase separation) to promote crystal nucleation, a smaller region with lower Ge content is needed if the critical nucleus is smaller, which would reduce the  time needed for the sizable fluctuation in the local Ge content to occur, and consequently also the nucleation time is reduced. 
  The crystal growth velocity v$_g$ at 500 K for this system (phase separated at 600 K and then quenched at 500 K) is shown in Table \ref{velocities}.
 We remark that the a-Ge regions remain amorphous after crystallization of GeTe at 500 and 600 K. The same crystallization behavior  of the  system phase separated at 600 K and then quenched at 500 K, was observed in the second simulation with an independent model mentioned above.
 We also repeated a simulation at 500 K for a homogeneous system (19200 atom) with composition Ge$_{53}$Te$_{47}$ generated by quenching from the melt and we similarly observed the formation of 5 crystalline nuclei after 0.5 ns. The crystal growth velocity is the same of that obtained for the phase separated system, as given in Table \ref{velocities}.

Finally, to assess if the system might undergo phase separation also at lower temperatures, we quenched the liquid model from 1200 K to  500 K in 70 ps and we then  performed a NVT simulation at this temperature for 14 ns.  The percentage of segregated Ge atoms saturates around 6 \% after about 5 ns, as shown in Fig. \ref{segreg_snap}b and no crystallization was observed. 
We can conceive that the system does not segregate in Ge$_2$Te at 500 K due to the lower atomic mobility of Ge atoms at this temperature.  The diffusion coefficient as a function of temperature, shown in Fig. \ref{cgdiff}, features a super-Arrhenius behavior with a drop from $1.2\cdot10^{-5}$ cm$^2$/s at 600 K to $2\cdot10^{-6}$ cm$^2$/s at 500 K.
However, the slow down with temperature of the kinetics of  Ge segregation is not simply due to the decrease of the growth velocity of the a-Ge regions. 
Would this be the case, we should see  an increase  in the growth time by about a factor six at 500 K with respect to 600 K (actually cube root of six if we consider an Avrami theory with just the growth of pre-formed nuclei). 
In fact, while at 600 K we see the formation of several overcritical nuclei of a-Ge like  in the time span of few hundreds ps, at 500 K no overcritical nuclei of a-Ge form in 15 ns.
Therefore nucleation of a-Ge region seems to control the kinetic of phase separation.

\begin{figure}[htbp!]
    \centering
    \centerline{\includegraphics[width=0.4\textwidth]{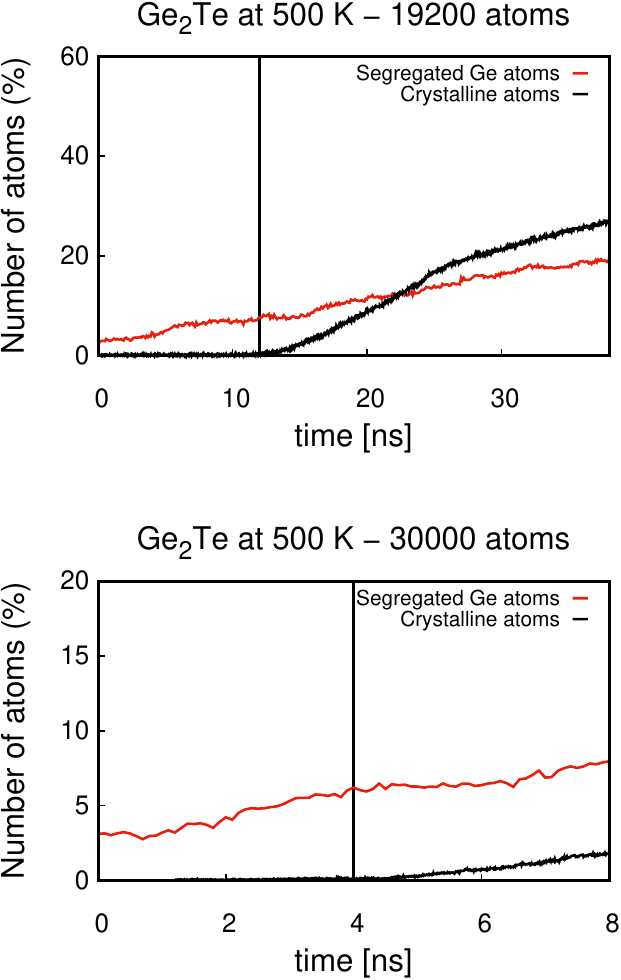}}
    \caption{Evolution in time of the number of crystalline atoms and of segregated Ge atoms  at 500 K in the simulation of (upper panel) the second 19200-atom  model and (lower panel) the larger 30000-atom model. In both models  crystal nucleation occurs before phase separation.}
    \label{500secondsmall}
\end{figure}

However, when we repeated the simulation at 500 K for the second independent model (19200 atoms) generated in the same manner of the first, a surprise came. 
After 12 ns, we observed the formation of a single overcritical nucleus of crystalline GeTe that
grew very slowly by expelling the excess of Ge.
Surprisingly, crystal nucleation occurred without prior phase segregation in a region where the local content of Ge-Ge bonds is  lower than the average in the Ge$_2$Te homogeneous model and also lower than the average of homogeneous  GeTe.
A snapshot of the distribution of Ge atoms with a fraction of Ge-Ge bonds lower than the average in a-GeTe (0.79) is shown in Fig. S8 in the Supplementary Information.
The crystal growth velocity is  low as shown in Table \ref{velocities} because of the need of expel a large amount of Ge. The evolution in time of the number of crystalline atoms and of segregated Ge atoms in the simulation at 500 K of the second 19200-atom model are compared in Fig. \ref{500secondsmall}.
A similar behaviour was observed for a larger 30000-atom model
quenched from the melt at 500 K in which crystal nucleation occurs before phase separation, albeit after about 4 ns (Fig. \ref{500secondsmall}).

\begin{figure}[htbp!]
    \centering
    \centerline{\includegraphics[width=0.3\textwidth]{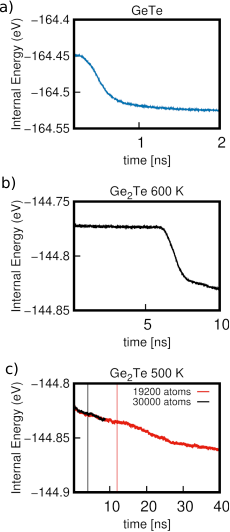}}
    \caption{Evolution in time of the internal energy per atom for a)
    stoichiometric GeTe at 600 K (32768-atom cell), b) Ge$_2$Te (30000-atom cell) at 600 K, c) Ge$_2$Te in the small (19200-atom, red line) and large (30000-atom, black line) models at 500 K. Vertical lines in panel c) indicate the onset of crystallization in the two models.}
    \label{internalenergy}
\end{figure}

The dependence of the nucleation time  on the different simulation conditions can be seen in Fig. \ref{internalenergy}  where the internal energy is reported as a function of time. A sudden drop of the internal energy corresponds to the onset of crystallization in GeTe and in Ge$_2$Te at 600 K. At 500 K the crystal growth is slow and the decrease in energy due to crystallization superimposes to the effect due to  equilibration of the supercooled liquid which is slow as well for the
off-stoichiometric system at this temperature.

We have also studied the crystallization at 500 K and 600 K of an amorphous 30000-atom model generated by quenching from 1200 K to 300 K in 100 ps and then annealed abruptly at the two target temperatures.
The results are similar to those of the liquid model supercooled from the melt discussed above.
The overheated amorphous phase crystallizes concurrently with the phase separation at 500 K, while phase segregation precedes crystal nucleation at 600 K.
The evolution in time of the fraction of segregated atoms and of the fraction of crystalline atoms for this last model at 500 and 600 K are shown in Fig. S9 in the 
Supplementary Information.

Therefore, we have seen two different regimes at different temperatures.
At high temperature (600 K), the nucleation time  of a-Ge is very short  and phase separation with segregation of most of Ge in excess occurs in a few ns. This yields a GeTe-like region with composition Ge$_{53}$Te$_{47}$ where the nucleation time for  crystalline GeTe is a few ns  in a 30000-atom cell, to be compared with the value of a few hundreds of ps in stoichiometric GeTe at the same temperature and for a similar cell size (32768 atoms). At 600 K, phase separation occurs first and GeTe crystallization follows in a two steps mechanism.
At lower temperatures (500 K), phase separation does not occur on the time scale of our simulations because the overcritical nuclei of a-Ge do not form, albeit Ge is still very mobile. 
Still, nucleation of crystalline GeTe is possible at 500 K without phase separation, but with a low crystal growth velocity because Ge in excess must be expelled during growth. At 500 K, crystallization is a single step process with Ge segregation occurring simultaneously with the growth of GeTe crystallites. 

 We attempt now to make a contact with the experimental results of Refs \cite{carria,gourvest,raoux}. The single step crystallization that we see at 500 K seems consistent with the crystallization seen in Refs. \cite{gourvest,raoux} at 470 K. 
 We must remark, however, that our models are generated
by quenching from the melt which corresponds to the operation conditions of the memory devices,  while  in Refs. \cite{gourvest,raoux} crystallization was studied experimentally in the asdep samples.
On the other hand, the two steps crystallization we see at 600 K, seems consistent with the two steps process (first segregation of Ge and then crystallization of GeTe)  seen in Ref  \cite{carria} at 600-630 K.
Our simulations can not, however, explain why a single step crystallization at a lower temperature about 500 K was not seen in Ref. \cite{carria}.  We can not exclude that our NN potential  could overestimate the nucleation rate of GeTe, as it seems to occur \cite{han2020} for the NN potential for stoichiometric GeTe that we developed previously
\cite{SossoNN}.

\section{Conclusions}
 In summary, we have generated a machine-learned potential for the Ge-rich Ge$_x$Te binary alloy by using the NN method implemented in the DeePMD package. We assessed that the NN potential can accurately reproduce the structural properties of the amorphous, liquid, and crystalline phases of Ge, GeTe, and Ge$_2$Te alloys. Then, this potential was used to study the crystallization mechanism of Ge$_2$Te by  MD simulations of 19200-atom and 30000-atom cells  lasting up to 40 ns. In stoichiometric GeTe, the system crystallizes in a few ns at both 600 and 500 K, consistently with previous simulations in literature \cite{SossoCryst}. In Ge$_2$Te instead,  we see a different behaviour at different temperatures. 
At 500 K, we saw in some models the formation of a crystalline nucleus of GeTe before  phase separation while in an other model (19200-atom cell)  crystal nucleation does not occur on our simulation time (14 ns). Nucleation starts in a point where the fraction of Ge-Ge bonds is lower than the average of stoichiometric a-GeTe. Crystal growth is, however, slow as all Ge in excess must be expelled during  growth. 
At this temperature, overcritical a-Ge nuclei do not form  on the time scale of 15 ns if not preempted by crystal  nucleation. On the contrary, at 600 K  phase separation takes place first with the formation of large and separated regions with average compositions Ge$_{88}$Te$_{12}$ and Ge$_{53}$Te$_{47}$. The slow down with temperature of the segregation process seems to be due to the increase of the  time needed for the formation of overcritical nuclei of a-Ge. 
Crystallization then occurs at 600 K in the Ge-poor region with a nucleation time still much longer than that of stoichiometric GeTe.

 Once a supercritical nucleus is formed in Ge$_{53}$Te$_{47}$ regions, crystal growth proceeds with a further expulsion of Ge in excess (because of a larger driving force for Ge segregation in the crystalline phase) which, however,  implies a still lower crystal growth velocity with respect to stoichiometric GeTe.  We remark that the a-Ge regions remain amorphous after crystallization of GeTe at 500 and 600 K on the time scale of our simulations.

Overall, our simulations represent a first step towards the atomistic modeling of the crystallization in Ge-rich GeSbTe alloys which are of interest for applications in embedded memories. Firstly, we have shown that it is possible to devise a reliable NN potential suitable to describe the full range of composition from Ge$_2$Te to GeTe and pure Ge. The database that we generated for this purpose will contribute to the training set for the development of a NN potential for the Ge-rich GeSbTe alloys which is in progress. Secondly, we have shown that phase separation with segregation of most of Ge in excess occurs on the relatively short time scale of a few ns
in Ge$_2$Te at high temperature (600 K). The complex process of Ge segregation and crystallization in non-stoichiometric alloys can thus be tackled by MD simulations.

\bibliography{biblio}

\medskip
\noindent
\textbf{Data availability} \par The NN potential,  the training DFT database, atomic trajectories of the phase separation and crystallization process, and a video of the relevant processes will be available in the Materials Cloud repository after acceptance of the article.
\medskip

\noindent
\textbf{Acknowledgements} \par 
The project has received funding from European
Union NextGenerationEU through the Italian Ministry of
University and Research under PNRR M4C2I1.4 ICSC 
Centro Nazionale di Ricerca in High Performance Computing,
Big Data and Quantum Computing (Grant No.
CN00000013). 
We acknowledge the CINECA award under the ISCRA initiative, for the availability of high-performance computing resources and support.

\medskip
\noindent
\textbf{Code availability} \par
LAMMPS, DeePMD and CP2k  are free and open source codes available at https:// lammps.sandia.gov,  http://www.deepmd.org and http://www.cp2k.org, respectively.

\medskip
\noindent
\textbf{Competing interests} \par
The authors declare no competing interests.

\medskip
\noindent
\textbf{Authors contributions} \par
D.B. and M.B. designed the research, performed the research and analyzed data. O.A.E contributed to the generation of the NN potential. D.B. and M.B. wrote the paper and all the authors edited the manuscript before submission.

\end{document}

\setcounter{figure}{0}
\begin{figure*}[htbp!]
    \centering
    \renewcommand\figurename{Figure~S$\!\!$}
    \includegraphics[width=0.6\textwidth]{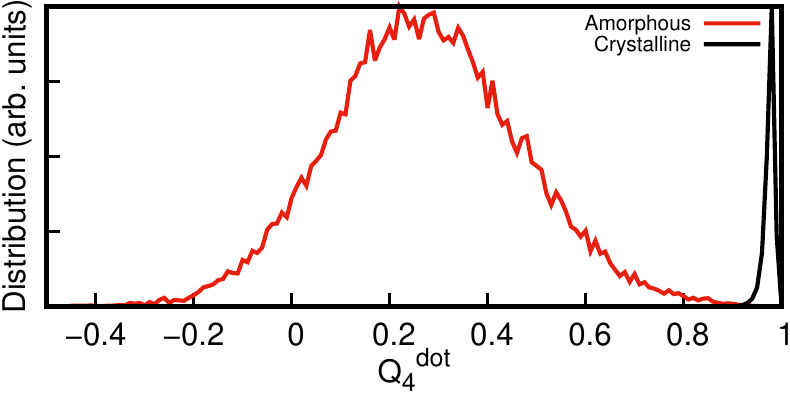}
    \caption{Distribution of the Q$^{dot}_{4}$ parameter in amorphous and crystalline cubic GeTe at 600 K in 4096-atom models.}
    \label{qdist}
\end{figure*}

\setcounter{figure}{1}
\begin{figure*}[htbp!]
    \centering 
    \renewcommand\figurename{Figure~S$\!\!$}
    \includegraphics[width=0.5\textwidth]{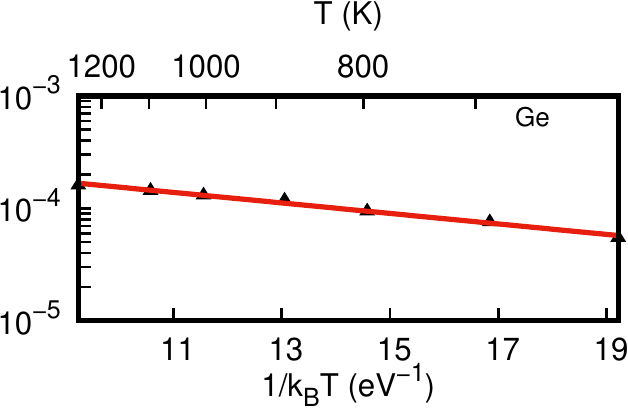}
    \caption{Diffusion coefficient ($D$) as a function of temperature of pure Ge from NN simulations. The experimental value for $D$ at 1273 K is 1.12 10$^{-4}$ cm$^2$/s (S. Chathoth {\sl et al.,}  Appl. Phys. Lett.  {\bf 94}, 221906 (2009)).}
    \label{cgdiffGe}
\end{figure*}

\setcounter{figure}{2}
\begin{figure*}[h]
    \centering
    \renewcommand\figurename{Figure~S$\!\!$}
    \includegraphics[width=0.6\textwidth]{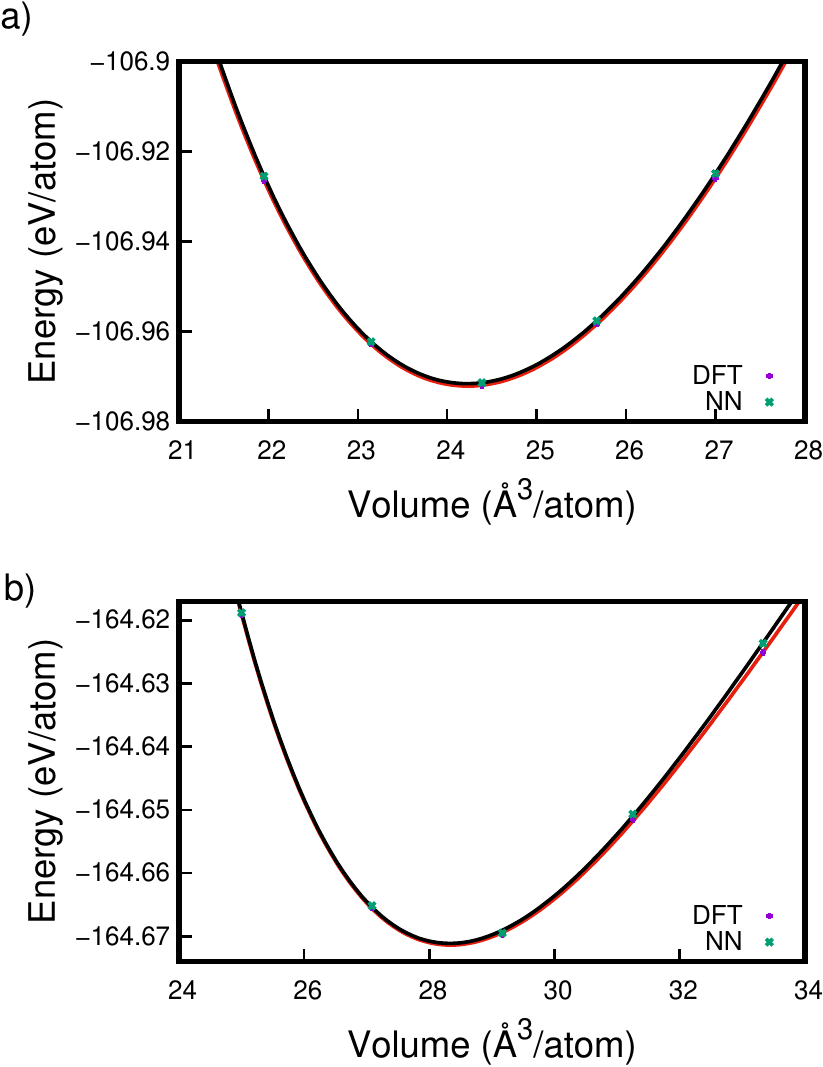}
    \caption{Equation of state from DFT and NN calculations of a) the cubic phase of Ge and b) the trigonal phase of GeTe. The continuous lines are Birch-Murnaghan fit  $E(V) = \frac{3B_0}{2} [ (V/V_0)^{7/3} -  (V/V_0)^{5/3} ]   \{ 1+ \frac{3}{4}(B'-4) [ (V/{V_0})^{2/3} -1 ]  $, where V$_0$ is the equilibrium volume, $E_0$ the equilibrium energy, $B_0$ the bulk modulus and $B^\prime$ its derivative respect to pressure. The fitting parameters are given in Table V in the article.}
    \label{BM_GE}
\end{figure*}

\setcounter{figure}{3}
\begin{figure*}[h]
    \centering
    \renewcommand\figurename{Figure~S$\!\!$}
     \includegraphics[width=0.42\textwidth]{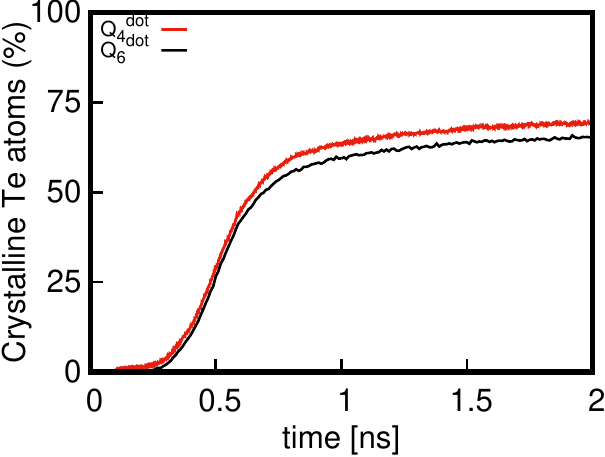}
    \caption{Evolution in time of the Q$^{dot}_{4}$ and Q$^{dot}_{6}$ crystalline order parameters in the crystallization process of GeTe at 600 K. The Q$^{dot}_{4}$ parameter measures the ordering of Te-Ge bonds (nearest neighbor distances in the simple cubic lattice neglecting chemical order) while the Q$^{dot}_{6}$ parameter measures the ordering of the Te fcc sublattice only (see article).}
    \label{Q6_Te}
\end{figure*}

\setcounter{figure}{4}
\begin{figure*}[h]
    \centering
    \renewcommand\figurename{Figure~S$\!\!$}
    \includegraphics[width=0.8\textwidth]{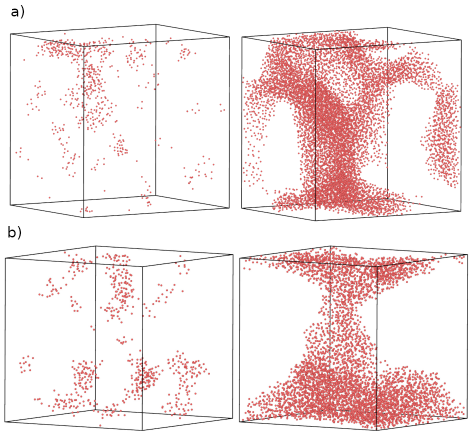}
    \caption{Snapshots of amorphous Ge-like atoms in the simulation at 600 K of  a) Ge$_2$Te  in the large 30000-atom model at the beginning (left) and after 10 ns  (right)  and
   b)  Ge$_2$Te  in a second  independent 19200-atom model at the beginning (left) and after 14 ns  (right). Only atoms with an environment close to that of a-Ge are shown (SOAP $k_j$ $>$ 0.92).}
    \label{segreg_snap2}
\end{figure*}

\setcounter{figure}{5}
\begin{figure*}[h]
   \centering
   \renewcommand\figurename{Figure~S$\!\!$}
    \includegraphics[width=0.6\textwidth]{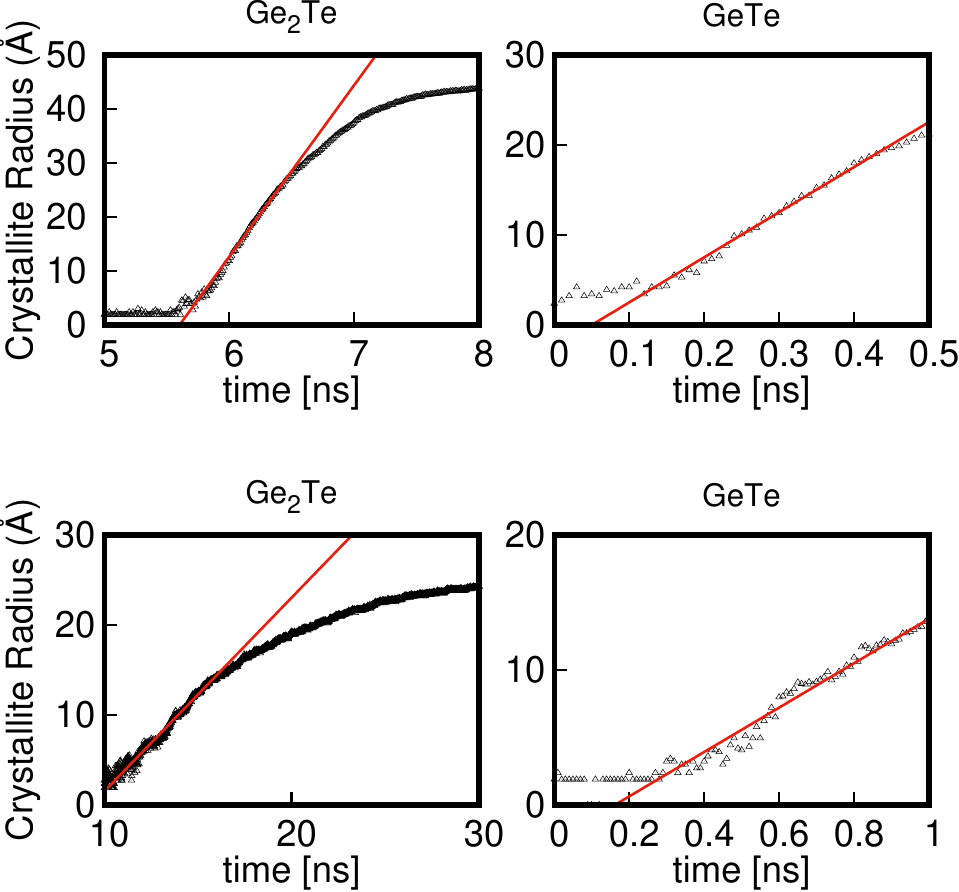}
 \caption{The evolution in time of the radius $R(t)$ of   a single GeTe crystalline nucleus in the simulation of (upper panels) stoichiometric GeTe and 
 partially phase separated Ge$_2$Te  at 600 K (30000-atom model, see article) and (lower panels) stoichiometric GeTe  and homogeneous (no phase separation, 19200-atom model) Ge$_2$Te  at 500 K.}
\end{figure*}

\setcounter{figure}{6}
\begin{figure*}[htbp!]
    \centering
    \renewcommand\figurename{Figure~S$\!\!$}
    \centerline{\includegraphics[width=1\textwidth]{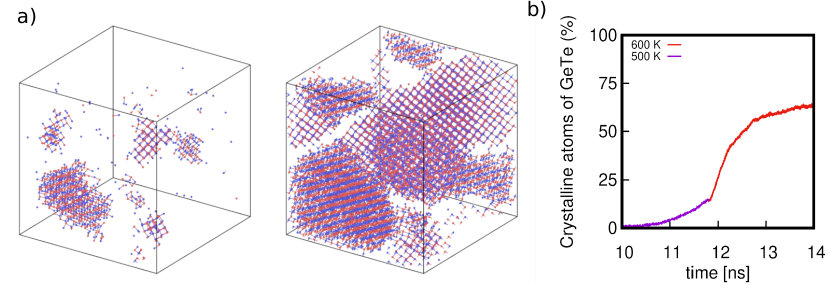}}
    \caption{ a) Crystalline atoms in the segregated 19200-atom model of Ge$_2$Te at t=1 ns after quenching to 500 K (left) and subsequent re-heating to 600 K (right) at t=4 ns. b) Fraction of crystalline atoms (defined by Q$^{dot}_{4}$)  as a function of time in the model of Ge$_2$Te with partial phase separation cooled at time t=10 ns from 600 K to 500 K, followed for 2.2 ns at 500 K   and then reheated to 600 K  to accelerate crystal growth.}
    \label{crystal_ge2teSmall}
\end{figure*}

\setcounter{figure}{7}
\begin{figure*}[h]
   \centering
   \renewcommand\figurename{Figure~S$\!\!$}
    \includegraphics[scale=1.5]{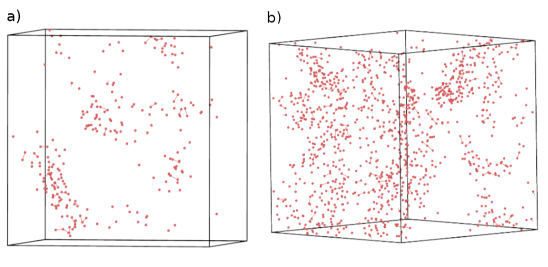}
 \caption{Snapshot of the distribution of Ge atoms with a fraction of Ge-Ge bonds lower than the average value in amorphous GeTe for  a) the second 19200-atom model at 500 K and b) the first 19200-atom model first phase separated at 600 K and then quenched at 500 K. In both models, crystal nucleation occurs in the regions highlighted in these snapshots.}
\end{figure*}

\setcounter{figure}{8}
\begin{figure}[htbp!]
    \centering
       \renewcommand\figurename{Figure~S$\!\!$}
    \centerline{\includegraphics[width=0.8\textwidth]{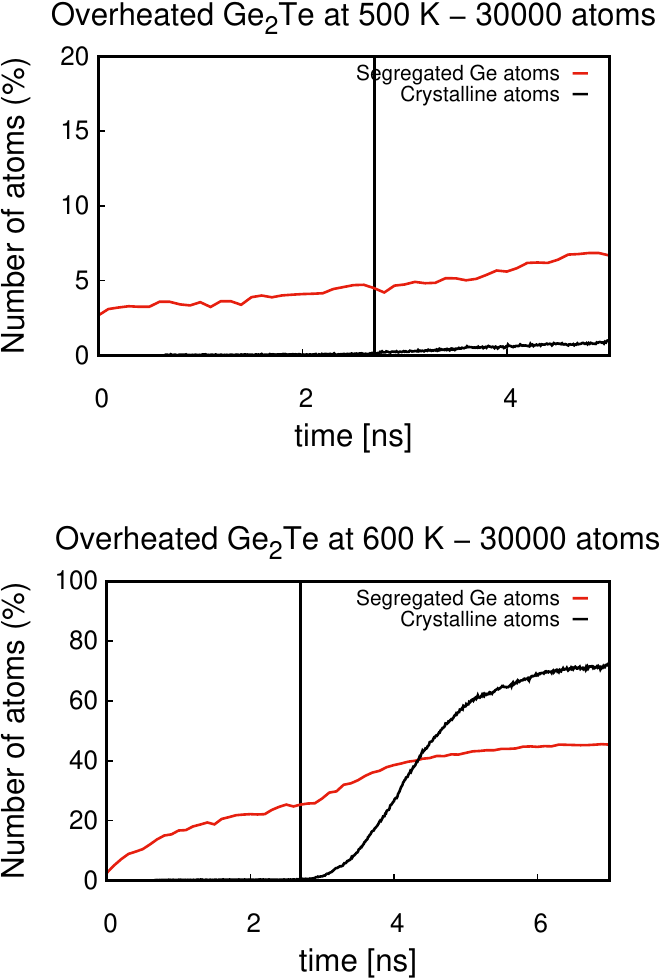}}
    \caption{Evolution in time of the number of crystalline atoms and of segregated Ge atoms  at 500 K  (upper panel) and  at 600 K (lower panel) of a  30000-atom  model of amorphous Ge$_2$Te generated by quenching from the melt to 300 K and then annealed at the two target temperatures. The vertical lines indicate the onset of crystallization}
    \label{amorphousoverheated}
\end{figure}
\vfill